\documentclass[aps,prl,twocolumn,showpacs]{revtex4}

\usepackage{epsfig}
\usepackage{graphicx}

\DeclareGraphicsExtensions{.eps,.EPS}

\begin{document}

\title{Spontaneous demagnetization of a dipolar spinor Bose gas at ultra-low magnetic field}
\author{B. Pasquiou, E. Mar\'echal, G. Bismut, P. Pedri, L. Vernac, O. Gorceix and B. Laburthe-Tolra}
\affiliation{Laboratoire de Physique des Lasers, CNRS UMR 7538,
Universit\'e Paris 13, 99 Avenue J.-B. Cl\'ement, 93430
Villetaneuse, France}

\begin{abstract}

Quantum degenerate Bose gases with an internal degree of freedom, known as spinor condensates, are natural candidates to study the interplay between magnetism and superfluidity. In the spinor condensates made of alkali atoms studied so far, the spinor properties are set by contact interactions, while magnetization is dynamically frozen, due to small magnetic dipole-dipole interactions. Here, we study the spinor properties of S=3 $^{52}$Cr atoms, in which relatively strong dipole-dipole interactions allow changes in magnetization. We observe a phase transition between a ferromagnetic phase and an unpolarized phase when the magnetic field is quenched to an extremely low value, below which interactions overwhelm the linear Zeeman effect. The BEC magnetization changes due to magnetic dipole-dipole interactions that set the dynamics. Our work opens up the experimental study of quantum magnetism with free magnetization using ultra-cold atoms.

\end{abstract}


\date{\today}

\maketitle

Within optical dipole traps, it is possible to trap all Zeeman states of an atomic species in its electronic ground state. This has enabled the field of multi-component (spinor) Bose-Einstein condensates (BECs), $i.e.$ BECs with \textit{internal} degrees of freedom \cite{spinor,stenger98}. With short range interactions, collisions between atoms are described by various scattering lengths, which leads to spin-dependent contact interactions. New  phases arise, which have been investigated for $F=1$ and $F=2$ atoms, using Rb and Na, by studying miscibility \cite{stenger98}, spin dynamics \cite{Chang04,Schmaljohann04,Kronjager06,Black07}, and spin textures \cite{Sadler06}. In these experiments, spin-dependent contact interactions are extremely small compared to the linear Zeeman effect. In addition, the gas magnetization remains constant for all practical purposes because contact interactions are isotropic, and anisotropic dipole-dipole interactions between alkali atoms are negligible. Consequently, the true ground state of the system, with free magnetization, at extremely low magnetic fields has never been experimentally investigated.

The production of Cr BECs in optical dipole traps \cite{Griesmaier2005,BECCr} allows for the first time to study $S=3$ spinor physics, with a wealth of possible quantum nematic phases \cite{diener2006,santos2006}, depending on the different scattering lengths $a_{0,2,4,6}$ of the molecular potentials $S_m={0,2,4,6}$. Contrarily to the cases of Rb and Na where spin dependent interactions are very small due to similar scattering lengths in different molecular channels, Cr has large spin dependent contact interactions. One consequence is that one can reach the regime where these interactions overwhelm the Zeeman effect at experimentally accessible magnetic fields (up to 1 mG, compared to typically 10 $\mu$G for Rb). While large magnetic fields favor a ferromagnetic ground state, below a critical magnetic field depending on the interactions the Cr spinor ceases to be ferromagnetic \cite{diener2006,santos2006}, which we here observe.

An important feature of Cr atoms, arising from their large electronic spin, is the strength of long-ranged, anisotropic dipole-dipole interactions between them. These interactions are too weak to greatly modify the phase diagram of Cr at low magnetic fields \cite{diener2006,makela2007}, but they do crucially modify spinor properties. They introduce magnetization-changing collisions \cite{pasquiou2010}, allowing the study of spinor physics with free magnetization, with an  intriguing coupling between the spin degrees of freedom and mechanical rotation that resembles the Einstein-de-Haas effect \cite{santos2006,edhcoldatom1,edhcoldatom3,edhcoldatom4}. They are also expected to lead to  spin textures, as investigated in \cite{vengalattore2008}, similar to domain formation in ferro-magnets. Here, we study how dipolar interactions induce magnetization dynamics, as the magnetic field is quenched below an extremely low value, which corresponds to a phase transition separating two spinor phases of different magnetization.

\begin{figure}
\centering
\includegraphics[width= 2.2 in]{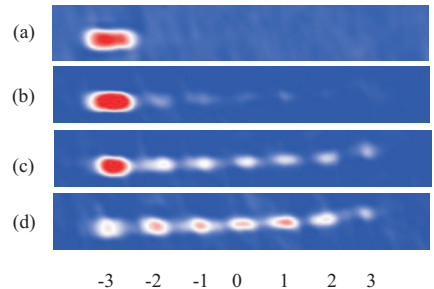}
\caption{\setlength{\baselineskip}{6pt} {\protect\scriptsize
Chromium BEC spin composition at low fields, revealed by Stern-Gerlach analysis (see Methods 1). The BEC spontaneously depolarizes as the magnetic field is lowered. Absorption pictures after 155 ms of hold time, in a field of: a) 1 mG; b) 0.5 mg c) 0.25 mG and d) 0 mG. At the lowest magnetic field, the spin distribution is $\left\{17.5 \pm 9,18 \pm 4,14 \pm 1.5, 15 \pm3, 17 \pm 3,12.5 \pm 4,6 \pm 2 \right\} \%$ corresponding to a magnetization of -0.5 $g_S \mu _B$. }}
\label{3ddep}
\end{figure}

In this article, we show that a Cr BEC in which atoms are polarized in the lowest single particle energy state $m_S=-3$ spontaneously depolarizes below a critical field of about 0.4 mG. We observe that the magnetic field below which depolarization occurs is larger if the peak local density of the BEC is increased by loading it into deep 2D optical lattices. This is a consequence of an increased meanfield energy, which raises the critical magnetic field up to 1.2 mG. We also analyze depolarization dynamics with and without the lattice, and the role of dipole-dipole interactions; we find that depolarization is slower in the lattice although the \textit{peak} density is then higher. This counterintuitive behavior is a consequence of a reduction of the \textit{average} density in the lattice and of the non-locality of the mean-field due to long-ranged dipolar interactions. In addition, without the lattices, we observe an increase of the thermal fraction, as expected from non-interacting spinor thermodynamics theory; however the observed depolarization of the BEC itself is a feature unique to the non trivial phase diagram at low magnetic field due to spin-dependent contact interactions.

We produce Cr BECs by performing evaporative cooling of atoms in the single particle lowest energy state $m_S=-3$, in an optical dipole trap \cite{BECCr}. The magnetic field during evaporation is sufficiently large that any two-body inelastic process is energetically forbidden. After a BEC has been obtained in $m_S=-3$, with typically 20 000 atoms, no discernible thermal fraction, and a peak density of $3.5 \times 10^{14}$ cm$^{-3}$ in an harmonic trap of frequencies $(320,400,550)$ Hz, we suddenly (within a $1/e$ time of 8 ms, set by the inductance of the coils and Eddy currents) reduce the magnetic field $B$ from 20 mG to an extremely low value, typically below 1 mG. 

The magnetic field is controlled by three orthogonal pairs of coils, and calibrated by rf spectroscopy. Extremely low magnetic fields correspond to magnetic resonance frequencies of the order of 1 kHz or less. To reduce technical noise, and provide better long term field stability, we have implemented an active stabilization of the ambient magnetic field (see Methods 1). With active stabilization, typical shot to shot magnetic field fluctuations are 100 $\mu$G.

If the final magnetic field is sufficiently small (below 0.5 mG typically), we observe a spontaneous depolarization of the BEC, while the total number of atoms remains constant. We measure the population in each of the Zeeman sublevels by performing a Stern-Gerlach experiment (see Methods 2). We finely tune the current sent to three orthogonal pairs of coils to maximize depolarization, which in practice precisely pinpoints $B=0$.  Typical depolarization results are displayed on Fig. \ref{3ddep}.

We have repeated the same experiment for a BEC adiabatically (within 15 ms) loaded in the ground state band of a 2D optical lattice \cite{Denschlag2002}, superimposed onto the optical dipole trap. Our lattice configuration is described in \cite{pasquiou2011}. The lattice depth in each direction is between 25 and 30 $E_R$, where $E_R = \frac{h^2}{2 m \lambda^2}$ is the recoil energy, with $\lambda=532$ nm, and $m$ the atom mass. The vibrational trapping frequency in a lattice site is 120 kHz, much larger than the chemical potential ($\mu/h=$11 kHz); consequently, the motion is frozen in two dimensions. We consider that the BEC is split into an array of 1D quantum gases, with an estimated peak density of $2 \times 10^{15}$ cm$^{-3}$.

\begin{figure}
\centering
\includegraphics[width= 2.8 in]{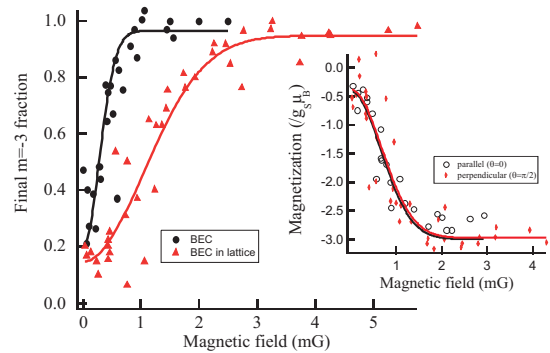}
\caption{\setlength{\baselineskip}{6pt} {\protect\scriptsize
Demagnetization curve for 3D BECs and for 1D quantum gases. The $m_S=-3$ final population is plotted as a function of the magnetic field. Full circles give the results for 3D-BECs and triangles for 1D tubes. Inset: final gas magnetization as a function of the magnetic field in the lattice, for two orthogonal orientations of the field with respect to axis of the lattice tubes. Lines are gaussian fits to the data.}}
\label{largeur}
\end{figure}

The fraction of atoms remaining in $m_S=-3$ after a hold time of 155 ms is displayed on Fig. \ref{largeur} for the two experimental configurations, as a function of magnetic field. While we observe strong depolarization at zero magnetic field for both configurations, the width of the depolarization curve is significantly larger for the BEC loaded in the 2D optical lattice than without the lattice. A gaussian fit to the experimental data gives a $1/e$ width of 0.4 mG for the BEC, and 1.2 mG for the BEC loaded in the lattice.

We interpret the increased width of the depolarization curve in the lattice as an effect of an increase of the mean-field interactions. Indeed, the chemical potential of the BEC is $\mu /h =$ 3.8 kHz, and reaches 11 kHz when loaded into the lattice. The widths of the depolarization curves are therefore roughly proportional to the chemical potential.

At $T=0$, and at finite magnetic field $B$, spontaneous depolarization from $m_S=-3$ becomes energetically possible when the energy cost to go from $m_S=-3$ to $m_S=-2$ is compensated for by a reduction of interaction energy between the atoms. In the case of Cr, because $a_4 < a_6$, this happens below a threshold $B_c$ \cite{diener2006}:
\begin{equation}
g_S \mu_B B_c \approx 0.7 \frac{2 \pi \hbar^2 (a_6-a_4) n}{m}
\label{Bc}
\end{equation}
with $g_S \approx 2$ the Land\'{e} factor of Cr atoms, $\mu_B$ the Bohr magneton and $n$ the density. $B_c$ corresponds to a critical magnetic field separating the ferromagnetic phase to either a polar phase (spin state $(\alpha,0,0,0,0,0,\beta)$), a cyclic phase (spin state $(\alpha,0,0,0,0,\beta,0)$), or  a phase in which all Zeeman states are populated, depending on the unknown value of $a_0$ \cite{diener2006,santos2006}. Phase separation is also possible \cite{he2009}. Let us compare the numerical values for $B_c$ to the values at which we do observe depolarization: for the BEC and the lattice respectively, $B_c$ calculated from eq. (\ref{Bc}) is respectively 0.25 mG and 1.15 mG, which is comparable to the observed 0.4 mG and 1.2 mG $1/e$ widths. The slight discrepancy between experiment and theory is consistent with our magnetic field fluctuations. We interpret the spontaneous depolarization  as due to the phase transition at $B_c$. Our work therefore differs from \cite{depolarizationcooling}, where depolarization of a thermal Cr gas occurred because the thermal energy was converted into Zeeman energy through dipolar collisions between atoms. 

We emphasize that technical noise can not mimic the increase of the width of the depolarization curve as the atoms are loaded in the lattice. Indeed, as the Zeeman effect is purely linear for Cr, rf spectroscopy, and more generally, the coupling of atoms to time-dependent magnetic fields, are independent of mean-field interactions: in a linear spin system, a magnetic field does not couple two different molecular potentials. We have also verified by rf spectroscopy that the quadratic effect related to the tensorial light shift of Cr \cite{quad} in the lattice is not responsible for the increased width of the depolarization curve \cite{notequad}.

\begin{figure}
\centering
\includegraphics[width= 2.8 in]{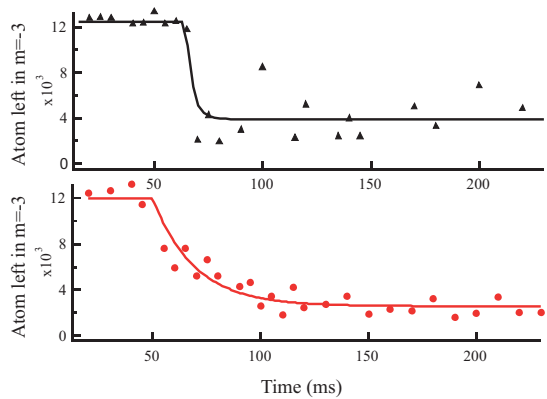}
\caption{\setlength{\baselineskip}{6pt} {\protect\scriptsize
 Time evolution of the $m_S=-3$ population, as a function of the time delay between the moment at which the magnetic field control is set to zero and the beginning of the Stern-Gerlach procedure. a) dynamics for 3D BECs; b) dynamics for 1D quantum gases. Full lines are guides to the eye. Depolarization starts after a delay, corresponding to the time for $B(t)$ to reach $B_c$ (see text).}}
\label{dynamique}
\end{figure}

We have investigated the dynamics of depolarization, with and without the lattice, as shown on Fig. \ref{dynamique}. At t=0, the magnetic field control is suddenly switched to 0. The magnetic field $B(t)$ does not instantaneously reach its final value, and in practice, for the first 50 ms, the atoms remain completely polarized as $B(t<50$ms$)>B_c$. Then, as $B(t)<B_c$, depolarization occurs. Without the lattice, depolarization occurs so suddenly ($\leq 5 $ ms) that we can barely resolve it; depolarization is slower in the lattice, although the peak density is then much larger. As shown below, the typical timescale for depolarization is set by the magnetic field resulting from the surrounding dipolar atoms, $i.e.$ by the non-local meanfield due to dipolar interactions \cite{edhcoldatom1}. For atoms loaded into an optical lattice, the large increase of the (repulsive) contact mean-field forces the cloud to swell in our experimental configuration; the overall volume of the cloud is then increased by a factor of about three, hence reducing the dipolar mean-field. A slower depolarization dynamics in the lattice is thus a consequence of the non local character of dipole-dipole interactions, and indicates inter-site inelastic dipolar couplings in the lattice.

Another signature for inter-site coupling is provided in the inset of Fig. \ref{largeur}, where we show that the depolarization is identical for two different orientations $\theta$ of the magnetic field, either parallel or perpendicular to the tubes axis. In a previous work, where we have studied dipolar relaxation in 2D optical lattices for atoms in $m_S=3$ at larger magnetic fields, the magnetization changing collisions occur in a given lattice site, and they are suppressed for $\theta=0$ because of cylindrical symmetry \cite{pasquiou2011}. The situation is different here because the experiment is performed at much smaller magnetic fields, and dipolar relaxation occurs at increasing inter-atomic distances for decreasing magnetic field \cite{pasquiou2010}. At 1 mG, the typical distance at which dipolar relaxation occurs is 300 nm, exceeding the lattice periodicity, which breaks cylindrical symmetry. A description including inter-sites coupling is hence required. 

To account for the depolarization dynamics, we consider the time-dependent Gross-Pitaevskii equation, starting with a polarized BEC in $m_S=-3$, including contact and dipole-dipole interactions. Dipole-dipole interactions provide the only mechanism to transfer the atoms from the maximally polarized $m_S=-3$ state to $m_S=-2$. We calculate the dynamics at short times by assuming that the population in $m_S=-2$ remains small, and that population in $m_S>-2$ is negligible. Then, the coupling between the fields $\phi_{j=(-3,-2)}$ describing the $m_S={(-3,-2)}$ components is given by (see also \cite{polonais}):

\begin{equation}
\Gamma(\vec{r})= \gamma \int d^3r' \frac{\left[(x-x')-i(y-y')\right](z-z')}{\left|\vec{r}-\vec{r'}\right|^5}\left|\phi_{-3}(\vec{r'})\right|^2
\label{evolution}
\end{equation}
where $\gamma= -3 S^{3/2} \hbar^2 d^2 / \sqrt{2}$, with $d^2=\mu_0 (g_S \mu_B)^2 /4 \pi$, and $\mu_0$ the magnetic constant. At short times, coupling between $m_S=-3$ and $m_S=-2$ atoms is only set by dipolar interactions via the term $\Gamma(\vec{r})$, and $\left|\Gamma(\vec{r})\right|^{-1}$ is the typical time-scale for depolarization at $B=0$. This timescale is about 3 ms  for the BEC case and 10 ms for the lattice case, in relatively good agreement with the observations (Fig. \ref{dynamique}) at the lowest achievable magnetic field.


\begin{figure}
\centering
\includegraphics[width= 2.8 in]{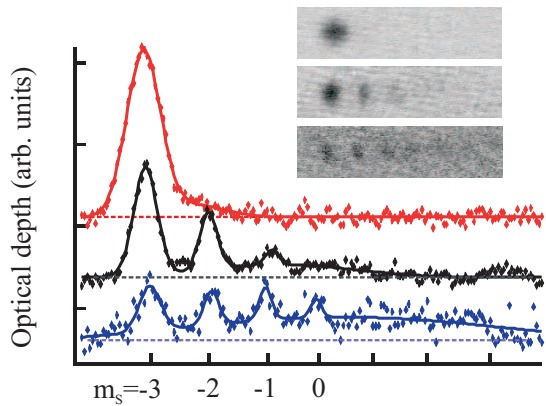}
\caption{\setlength{\baselineskip}{6pt} {\protect\scriptsize
Stern-Gerlach monitoring of the spin composition as the depolarization develops, for the BEC case. These pictures utilize our second Stern-Gerlach procedure (see Methods 2) which insures simultaneously spin separation and a measurement of the momentum distributions of each Zeeman components. The three curves, vertically offset for better clarity, correspond to the three absorption pictures shown in insert, taken at three steps of depolarization, separated by a few ms. Solid lines result from multiple gaussian fits. The width of the narrow peaks shows that, during depolarization, atoms in different spin states remain Bose condensed. A broad additional pedestal reveals the presence of a thermal component.}}
\label{SG}
\end{figure}

We have further monitored the state of the atoms at the first stages of depolarization. For this, we have implemented an alternative Stern-Gerlach procedure (see Methods 2), allowing a simultaneous analysis of spin and momentum distributions. As shown in Fig. \ref{SG}, soon after depolarization has started the momentum distributions of the Zeeman components remain extremely narrow, a signature that the system is still Bose condensed. These distributions are narrower than the one of the polarized BEC, presumably reflecting the decrease in mean-field energy in the depolarized case, since $a_6$ is the largest of the Cr scattering lengths. We could not detect vortices in the $m_S=-2$ component, although they are predicted, due to conservation of total angular momentum \cite{santos2006,edhcoldatom1}. It is not yet established whether the magnetic field is low enough for such vortices to be observed, nor how long they should survive in our non cylindrical trap in the presence of thermal excitations. This will be further investigated in future experiments.

By measuring the momentum distribution using standard time-of-flight techniques without separating the Zeeman components, we also observe a decrease of the condensate fraction when depolarization is important, although the temperature (deduced by fitting the thermal wings) does not change. Such condensate fraction reduction at constant temperature is a direct consequence of the release in the constraint of the spin degrees of freedom, which lowers the critical degeneracy temperature $T_c$ \cite{isoshima2000}. Although the reduction of $T_c$ might be understood in the framework of non-interacting spinor thermodynamics, it is worthwhile noting that depolarization of the BEC can not be understood in such framework, as a non-interacting BEC with free magnetization remains ferromagnetic for all magnetic fields \cite{simkin1999}.

In conclusion, this work represents the first investigation of quantum magnetism where magnetization is free and spin-dependent contact interactions set the many-body mean-field ground state. We have observed the magnetization dynamics of a BEC due to dipolar interactions, as the magnetic field is quenched through a critical value corresponding to (at $T$=0) a quantum phase transition. We have established that demagnetization is set by the dipolar mean-field. Demagnetization leads to a non-polarized BEC, although the spin state which is reached does not necessarily reveal the nature of the ground state, due to diabaticity at the phase transition, and to thermal excitations: temperature ($\approx 100$ nK) is comparable to the difference of energy between the spinor phases at $B=0$. 

We will next try to observe the rotation necessarily induced by demagnetization, in the spirit of the Einstein-de-Haas effect \cite{santos2006, edhcoldatom1, edhcoldatom3, edhcoldatom4}. We will also investigate how this rotating system thermalizes at low magnetic fields, to possibly reach quantum nematic (mean-field) phases \cite{diener2006}. When the BEC is loaded into the 2D lattice, we produce an array of mesoscopic samples (40 atoms per site) at low magnetic field, high density, and with large spin-dependent interactions; such conditions are known to be favorable to the observation of non-classical spin states \cite{bigelow} or fragmented BECs \cite{ho2000}.

This research was supported by the Minist\`ere de l'Enseignement Sup\'erieur et de la Recherche (within CPER) and by IFRAF.

\section{Methods}

\subsection{Magnetic field control}

The three components of the magnetic field along orthogonal directions of space are stabilized using an active cancellation of the AC and DC magnetic field fluctuations:  three large, 3-turn rectangular coils, with a typical size of $1.5 \times 1.7$ m, and with their center located about 1 m away from the BEC,  along the 3 spatial directions, are used as compensation coils. The magnetic field close to the experimental chamber is measured using a 3 axis fluxgate magnetic sensor (Bartington Mag 03MC). The sensor is located outside the experimental chamber, about 15 cm away from the BEC. The three components of the magnetic field are compared to a computer controlled set-value, with a resolution of 50 $\mu$G. The error signal is sent to a feedback loop which is a simple proportional controller, and the current in each coil is regulated with a 1 Amp push-pull type voltage-controlled current amplifier.

Without active compensation, the characteristics of the magnetic noise are the following: the main noise contribution comes from the 50 Hz AC noise of magnitudes, up to 4 mG peak/peak in one of the directions. This noise originates from the different equipment located around the experiment. The DC magnetic field fluctuates up to 2 mG during the day time. This is much larger than the natural fluctuations of the earth magnetic field which are of the order of 100 to 200 $\mu$G. These fluctuations are mainly caused by activities in the laboratory (positions of metallic objects) and outside the laboratory (elevators, cars).

When the active stabilization is switched on, the AC noise is decreased below 100 $\mu$G peak/peak and the DC magnetic field value is stabilized to better than 20 $\mu$G at the position of the sensor. We have also measured the residual magnetic field fluctuations with a second independent sensor located 20 cm away from the first one. We observe that the drifts of the DC magnetic field stay below 100 $\mu$G over 1 hour. At this position, though, the typical AC fluctuations are 500 $\mu$G peak/peak. This comes from the fact that some of the AC noise sources are not very far from the BEC (around 1 meter) so that the AC noise is not perfectly spatially homogeneous on a 20 cm scale. We have nevertheless checked by rf spectroscopy that this AC noise is efficiently screened by the metallic experimental chamber, so that the residual fluctuations at the BEC position are given by the 100 $\mu$G DC fluctuations. The experimental spectrum shown in Fig. \ref{largeur} corroborates this estimate.

\subsection{Stern and Gerlach procedures}

We have implemented two different Stern-Gerlach procedures to monitor the spin populations at low magnetic fields. For both procedures, a small (10 mG) magnetic field $B_{o}$ is first adiabatically applied to the atoms, with a rise time of 10 ms.  We insure that $B_{o}$ is large enough that a magnetic field gradient can also then be applied to the atoms to separate the Zeeman components adiabatically, $i.e.$ without changing the spin composition.

In the first Stern-Gerlach procedure, the applied magnetic field gradient is small (0.25 G/cm), and the magnetic field felt by each Zeeman component remains always small. The Zeeman components then separate very slowly. To prevent them from falling and keep them in the imaging field of view, their horizontal motion is channeled in one direction by the horizontal laser beam used for producing and trapping the BEC, and only the vertical beam is removed during the procedure. An absorption image is taken after 45 ms of 1D expansion in the magnetic field gradient, right after the horizontal optical trap is switched off. Due to the very small magnetic field employed, absorption imaging offers a good absolute determination of the number of atoms in each Zeeman component, as the Zeeman shifts are small compared to the transition linewidth for absorption imaging. We verify this by the fact that the sum of the populations in all Zeeman components is constant while the gas depolarizes. Unfortunately, the momentum distribution of each Zeeman component is lost when using this procedure, because the atoms have then moved out of the focus of the imaging system. This first procedure was used for all results in this paper, except the ones shown in Fig. \ref{SG}.

A second Stern-Gerlach procedure was used for Fig. \ref{SG}. We apply a stronger magnetic field gradient (1 G/cm). The Zeeman components separate sufficiently rapidly that the optical trap can be released almost simultaneously. The Zeeman components are separated in 5 ms, and this faster procedure also gives access to the momentum distribution of each Zeeman component, although the actual determination of their relative atom number is not as good as the one in the first procedure, because of the presence of a relatively strong magnetic field (a few G) when absorption imaging is performed.


\begin{thebibliography}{99}

\bibitem{spinor} T.-L. Ho, Phys. Rev. Lett. \textbf{81}, 742 (1998); T. Ohmi and K. Machida, J. Phys. Soc. Jpn. \textbf{67}, 1822 (1998).

\bibitem{stenger98} Stenger, J. et al. Nature 396, 345 (1998).

\bibitem{Chang04} M.-S. Chang, C. D. Hamley, M. D. Barrett, J. A. Sauer, K. M. Fortier, W. Zhang, L. You, and M. S. Chapman, Phys. Rev. Lett. \textbf{92}, 140403 (2004)

\bibitem{Schmaljohann04} H. Schmaljohann, M. Erhard, J. Kronjager, M. Kottke, S. van Staa, L. Cacciapuoti, J. J. Arlt, K. Bongs, and K. Sengstock,  Phys. Rev. Lett. \textbf{92}, 040402 (2004)

\bibitem{Kronjager06} J. Kronjager et al., Phys. Rev. Lett. \textbf{97}, 110404 (2006).

\bibitem{Black07} A. T. Black, E. Gomez, L. D. Turner, S. Jung, and P. D. Lett, Phys. Rev. Lett. \textbf{99}, 070403 (2007)

\bibitem{Sadler06} L. E. Sadler et al., Nature (London), \textbf{443}, 312 (2006)

\bibitem{Griesmaier2005} Axel Griesmaier, Jorg Werner, Sven Hensler, Jurgen Stuhler, and Tilman Pfau, Phys. Rev. Lett. \textbf{94}, 160401 (2005)

\bibitem{BECCr} Q. Beaufils, R. Chicireanu, T. Zanon, B. Laburthe-Tolra, E. Marechal, L. Vernac, J.-C. Keller, and O. Gorceix, Phys. Rev. A \textbf{77}, 061601 (2008) G. Bismut, B. Pasquiou, D. Ciampini, B. Laburthe-Tolra and E. Marechal, L. Vernac, and O. Gorceix. Applied Physics B, \textbf{102}, 1 (2011)

\bibitem{diener2006} Roberto B. Diener and Tin-Lun Ho, Phys. Rev. Lett. \textbf{96}, 190405 (2006)

\bibitem{santos2006} L. Santos and T. Pfau, Phys. Rev. Lett. \textbf{96}, 190404 (2006)

\bibitem{makela2007} H. Makela and K.-A. Suominen, Phys. Rev. A \textbf{75}, 033610 (2007) 

\bibitem{pasquiou2010} B. Pasquiou, G. Bismut, Q. Beaufils, A. Crubellier, E. Marechal, P. Pedri, L. Vernac, O. Gorceix, and B. Laburthe-Tolra, Phys. Rev. A \textbf{81}, 042716 (2010)

\bibitem{edhcoldatom1} Yuki Kawaguchi, Hiroki Saito, and Masahito Ueda, Phys. Rev. Lett. \textbf{96}, 080405 (2006)

\bibitem{edhcoldatom3} Krzysztof Gawryluk, Miroslaw Brewczyk, Kai Bongs, and Mariusz Gajda, Phys. Rev. Lett. \textbf{99}, 130401 (2007)

\bibitem{edhcoldatom4} B. Sun and L. You, Phys. Rev. Lett. \textbf{99}, 150402 (2007)

\bibitem{vengalattore2008} M. Vengalattore, S. R. Leslie, J. Guzman, and D. M. Stamper-Kurn, Phys. Rev. Lett. \textbf{100}, 170403 (2008)

\bibitem{Denschlag2002} Denschlag JH, Simsarian JE, Haffner H, et al. JOSA B, \textbf{35}, 3095 (2002) 

\bibitem{pasquiou2011} B. Pasquiou, G. Bismut, E. Marechal, P. Pedri, L. Vernac, O. Gorceix, and B. Laburthe-Tolra, Phys. Rev. Lett. \textbf{106}, 015301 (2011)

\bibitem{he2009} Liang He and Su Yi, Phys. Rev. A \textbf{80}, 033618 (2009) 

\bibitem{depolarizationcooling} M. Fattori, T. Koch, S. Goetz, A. Griesmaier, S. Hensler, J. Stuhler, T. Pfau, Nature Physics 2, 765 (2006)

\bibitem{quad} L. Santos, M. Fattori, J. Stuhler, and T. Pfau, Phys. Rev. A \textbf{75}, 053606 (2007), Chicireanu R, Beaufils Q, Pouderous A, et al. Eur. Phys. J D \textbf{45} 189 (2007)

\bibitem{notequad} To rule out an impact of the quadratic effect, we have performed rf spectroscopy in the optical lattice. We could not observe any quadratic shift of the rf resonance frequency, which sets an upper bound for the tensorial light shift  of 300 Hz $\times m_S^2$, insufficient to account for the 1.2 mG width of the depolarization curve in the lattice. At this level, though, the quadratic effect may impact the nature of the ground state at low magnetic fields, see L. Santos, M. Fattori, J. Stuhler, and T. Pfau, Phys. Rev. A \textbf{75}, 053606 (2007).

\bibitem{polonais} see T. Swislocki et al.,  arXiv:1102.1566 (2011).

\bibitem{isoshima2000} T. Isoshima, T. Ohmi and K. Machida, J. Phys. Soc. Jpn., \textbf{69}, 3864 (2000)

\bibitem{simkin1999} M. V. Simkin and E. G. D. Cohen, Phys. Rev. A \textbf{59}, 1528 (1999)

\bibitem{bigelow} C. K. Law, H. Pu, and N. P. Bigelow, Phys. Rev. Lett. \textbf{81}, 5257 (1998)

\bibitem{ho2000} Tin-Lun Ho and Sung Kit Yip, Phys. Rev. Lett. \textbf{84}, 4031 (2000)


\end{thebibliography}
\end{document}